\documentclass[pdflatex, sn-basic, Numbered]{sn-jnl}

\usepackage{amsmath}

\usepackage{xurl}

\usepackage{subcaption}
\usepackage{tabularx}
\usepackage{booktabs}
\usepackage{xcolor}
\usepackage{manyfoot}

\usepackage{graphicx}
\graphicspath{ {figures/} }

\begin{document}

\title{A Graphical Model of the False Information and Fact-Checking Ecosystem}

\author[1]{\fnm{Haiyue} \sur{Yuan}}
\email{h.yuan-221@kent.ac.uk}
\equalcont{These authors contributed equally to this work.}
\author[2]{\fnm{Enes} \sur{Altuncu}}
\email{enes.altuncu@iuc.edu.tr}
\equalcont{These authors contributed equally to this work. The study was conducted while the second author was a PhD student at the University of Kent.}
\author*[1]{\fnm{Shujun} \sur{Li}}
\email{S.J.Li@kent.ac.uk}
\author[3]{\fnm{Can} \sur{Ba\c{s}kent}}
\email{c.baskent@mdx.ac.uk}
\author[1]{\fnm{Jason R.C.} \sur{Nurse}}
\email{J.R.C.Nurse@kent.ac.uk}

\affil[1]{\orgdiv{Institute of Cyber Security for Society (iCSS) \& School of Computing}, \orgname{University of Kent}, \city{Canterbury}, \postcode{CT2 7FS}, \country{UK}}
\affil[2]{\orgdiv{Department of Computer Engineering}, \orgname{Istanbul University-Cerrahpaşa}, \city{Istanbul}, \postcode{34320}, \country{Türkiye}}
\affil[3]{\orgdiv{Department of Computer Science}, \orgname{Middlesex University}, \city{London}, \postcode{NW4 4BT}, \country{UK}}

\abstract{The wide spread of false information online, including misinformation and disinformation, has become a major problem for our highly digitised and globalised society. A lot of research has been done to better understand different aspects of false information online such as behaviours of different actors and patterns of spreading, and also on better detection and prevention of such information using technical and socio-technical means. One major approach to detecting and debunking false information online is to use fact-checkers. Despite a lot of research on online false information, we noticed a lack of conceptual models describing the complicated ecosystem of false information and fact-checking. In this paper, we introduce the first comprehensive graphical model of the ecosystem, focusing on false information online in multiple contexts, including traditional media outlets, as well as user-generated and AI-generated content. The proposed enhanced entity-relationship (EER) model covers a wide range of entities and relationships, and it can be a new useful tool for researchers and practitioners to study false information online and the effects of fact-checking. To demonstrate its usefulness, we provide two example applications of our proposed model, modelling real-world scenarios and reviewing the research literature.}


\maketitle

\section{Introduction}

In today's highly digitised and globalised society, the wide propagation of false information has become a more serious problem than ever. Different types of false information, including misinformation and disinformation~\citep{Howard-P2021, Bin-G2020, Wardle-C2017}, are among the sources of different forms of online harms on various online platforms such as social media, instant messaging applications, web forums, online chat rooms, and online news portals. According to the Global Risks Report published by the World Economic Forum in 2025, false information is expected to become the most severe threat in the next two years~\citep{wef2025}. Many researchers have studied different aspects of online false information, which include how different actors behave when facing false information, how false information propagates, and how such harmful information can be detected and mitigated via technical and socio-technical means~\citep{guo2020, kumar2018}.

The use of human, organisational, and automated fact-checkers is a common way to detecting and debunking online false information. Many fact-checking organisations and networks have been established, such as Full Fact in the UK\footnote{\url{https://fullfact.org/}} and PolitiFact in the US\footnote{\url{https://www.politifact.com/}}. Although many researchers have investigated fact-checking~\citep{nakov2021, juneja2022}, to the best of our knowledge, there is a lack of conceptual models that describe the complicated ecosystem of false information online and fact-checkers. Such conceptual models can help develop a better understanding of different stakeholders, relationships and processes that are involved in the ecosystem and enable more comprehensive analyses of the problem of false information at the ecosystem level. To fill this research gap, in this work, we propose the first comprehensive graphical model of the ecosystem mentioned above, focusing on modelling false information online in multiple contexts, including traditional media outlets, user-generated and AI-generated content~\citep{park2025}, and different types of fact-checkers. The proposed enhanced entity-relationship (EER) model contains a wide range of entities and relationships, and we provide two example applications, i.e., modelling real-world scenarios and reviewing the research literature, to show the usefulness of the proposed model.

The rest of the paper is organised as follows. Section~\ref{sec:related_work} presents related work on false information and fact-checking. Section~\ref{sec:graphical_model} describes the proposed model of the false information and fact-checking ecosystem. Then, Section~\ref{sec:real-world_scenarios} and Section~\ref{sec:literature_analysis} demonstrate how the model can be used to describe different entities and relationships in two example applications -- modelling a couple of real-world scenarios and reviewing recent literature, respectively. Further discussions and future work are given in Section~\ref{sec:discussion}, and Section~\ref{sec:conclusion} concludes this paper.

\section{Related Work}
\label{sec:related_work}

\subsection{False Information Types}

Research has been conducted to study and classify false information to better understand its definition and scope. One popular approach is to categorise false information into misinformation and disinformation~\citep{Howard-P2021, Bin-G2020, Wardle-C2017}, where the discrimination between the two is subtle~\citep{Stahl-BC2006}. They can be distinguished by the creator's `intention'~\citep{Bin-G2020}. Thereby, misinformation can be interpreted as false information whose creator did not have the intention to cause harm~\citep{Scheufele-DA2019, Kumar-K2014}, e.g., the information is incorrect by accident. On the other hand, disinformation is about information that was intentionally created to cause harm via different approaches such as deceptive advertising, government propaganda, and doctored photographs~\citep{Fallis-D2015}. More specifically, the United Nations proposed its definition of misinformation as `the accidental spread of inaccurate information' while disinformation has been defined as information `not only inaccurate, but intends to deceive and is spread in order to do serious harm'\footnote{\url{https://www.un.org/en/countering-disinformation}}. In addition, \citet{Wardle-C2017} suggested an additional category of false information, \textit{malinformation} to describe the scenario when genuine information is shared to harm rather than serve the public interest, such as `true information that violates a person's privacy without public interest justification'~\citep{Wardle-C2018}.

Moreover, researchers have attempted to categorise false information from real-life scenarios, aiming to have a better understanding of the actors, the motivations, and how false information spreads. \citet{Zannettou-S2019} categorised false information on the web into eight types, including fabricated information, propaganda, conspiracy theories, hoaxes, biased or one-sided information, rumour, clickbait, and satire news. Moreover, they identified a number of false information actors (e.g., bots, criminal/terrorist organisations, governments, activist or political organisations, journalists) with different motives (e.g., malicious intent, influence, profit, passion).

Similarly, \citet{Meel-P2020} suggested that false information can be considered as information pollution with a number of identified motivations (e.g., political intent, financial profit, passion, increase customer base), and it has many formats, including rumour, fake news, misinformation, disinformation, clickbait, hoax, satire/parody, opinion spam, propaganda, and conspiracy theory.

To better explain the complexity of false information categorisation, \citet{Meel-P2020} argued that all the categories of false information share some heterogeneity. One example is fake news, which overlaps with satire, misinformation, rumour, disinformation, and opinion spam. Considering the dynamic context of fake news and its broad scope, \citet{Molina-D2021} further suggested that fake news is beyond the spectrum of false information and concluded that fake news should be considered as an umbrella term with seven categories including false information (false news), polarised content, satire, misreporting, commentary, persuasive information, and citizen journalism.

\subsection{Fact-checking}

Due to the negative impact of false information/fake news on different aspects of society at a global scale, academia, industry, and international and national organisations have made significant efforts to act against the generation and spread of false information. At the international level, the European Commission launched the major counter-disinformation initiatives in 2018~\citep{EU_Report2018a, Renda-A2018}, set up a high-level group of experts (HLEG) to advise on relevant policy initiatives, and introduced a Eurobarometer survey and self-regulatory Code of Practices for the industry~\citep{EU_Report2018b}.

There are many different fact-checking outlets across the globe, and all share the same goal of promoting truth to the public. \citet{Graves-L2016} studied fact-checking sites in Europe and suggested that fact-checkers in Europe can be categorised into three types: 1) \textit{reporters} normally consider the fact-checking mission in journalistic terms to inform the public; 2) \textit{reformers} primarily use fact-checking as part of an agenda of political reform; 3) \textit{experts} are more or less considered as a think tank.

In addition, \citet{Graves-L2016} argued that there are two main fact-checking models: 1) the \textit{newsroom model} is associated with established media companies, where the legacy news media remain the main source of political fact-checking, such as `FactCheck' program from Channel 4 News, `Reality Check blog' from Guardian; 2) the \textit{non-government organisation} (NGO) model represents fact-checking outlets operate independently or backed by NGOs or charities such as FullFact in the UK and Teyit\footnote{\url{https://teyit.org/}} in Türkiye.

World widely, the international fact-checking network (IFCN)\footnote{\url{https://www.poynter.org/ifcn/}} at Poynter\footnote{The Poynter Institute for Media Studies is a non-profit journalism school and research organisation in US \url{https://www.poynter.org/}.} has a growing community of fact-checking/fact-checker organisations to advocate of factual information in the fight against misinformation/disinformation/fake news, where verified signatories of the IFCN includes FullFact, PolitiFact and Teyit mentioned before, and many other organisations such as FactCheck\footnote{\url{https://www.factcheck.org/}} in the US, `20 Minutes Fake Off'\footnote{\url{https://www.20minutes.fr/}} in France, Maldita\footnote{\url{https://maldita.es/}} in Spain, Africa Check\footnote{\url{https://africacheck.org/fact-checks}} in South Africa, Annie Lab\footnote{\url{https://annielab.org/}} in Hong Kong, Australian Associated Press FactCheck\footnote{\url{https://www.aap.com.au/factcheck/}} in Australia, and BOOM\footnote{\url{https://www.boomlive.in/}} in India\footnote{The full up-to-date list can be found at \url{https://ifcncodeofprinciples.poynter.org/signatories}.}.

By reviewing the fact-checking and false information/fake news detection methodologies, \citet{Collins-B2021} classified them into multiple common approaches: experts/professionals fact-checker approach, crowd-sourced approach, machine learning approach, natural language processing technique, hybrid technique, expert-crowdsource approach, human-machine approach, graph-based method, deep learning approach, and recommendation system approach. Based on these different approaches, researchers have also developed fact-checking tools such as TweetCred~\citep{Gupta-A2014}, Hoaxy~\citep{Shao-C2016}, ClaimBuster~\citep{Hassan-N2017}, Fake News Tracker~\citep{Shu-K2019}, and NudgeCred~\citep{Bhuiyan-M2021}. In addition to the effort from researchers, players from the technology industry have also joined the battle of false information/fake news and developed many tools, such as the Fact Check Explorer developed by Google\footnote{\url{https://toolbox.google.com/factcheck/explorer}}, the Third-party fact-checking program of Meta\footnote{\url{https://www.facebook.com/journalismproject/programs/third-party-fact-checking}}, Twitter's Birdwatch platform\footnote{\url{https://twitter.github.io/birdwatch/}}, and AI-powered fact-checking platforms and services provided by Logically\footnote{\url{https://www.logically.ai}} and CheckStep\footnote{\url{https://www.checkstep.com}}.

\subsection{False Information and Fact-checking Ecosystem}
\label{subsec:fifc_ecosystem}

Despite a plethora of research on false information and fact-checking, less research has been done on the relationship between fact-checking (fact-checker) and false information and the associated models and ecosystems. This subsection overviews some related work we are aware of.

\citet{Jiang-S2018} investigated the linguistic signals expressed in social media comments in the presence of misinformation and fact-checking and found that fact-checking can have positive effects on linguistic signals but can also cause potential `backfire' effects. \citet{Burel-G2020} studied the relationship between misinformation and the spread of fact-checking information and reports during the COVID-19 pandemic. They observed the similarity in how both spread in social media and discovered that fact-checking information has a positive effect on combating misinformation but with a short span due to the overwhelming amount of misinformation spread compared with the smaller amount of fact-checking information.

In a similar study, \citet{Nicole-K2020} considered fact-checking for COVID-19 misinformation as risk communication and identified that fact-checking COVID-19 misinformation will be a difficult task given the complicated relationship and varying levels of trust. Furthermore, \citet{Yang-A2021} looked at the relationship between fact-checkers and spreaders of COVID-19 vaccine misinformation on Facebook. They revealed that fact-checkers' posts can receive more comments, however, misinformation spreaders can more easily use the URL co-sharing network to coordinate communication strategies.

\begin{figure}[!ht]
\centering
\includegraphics[width=\linewidth]{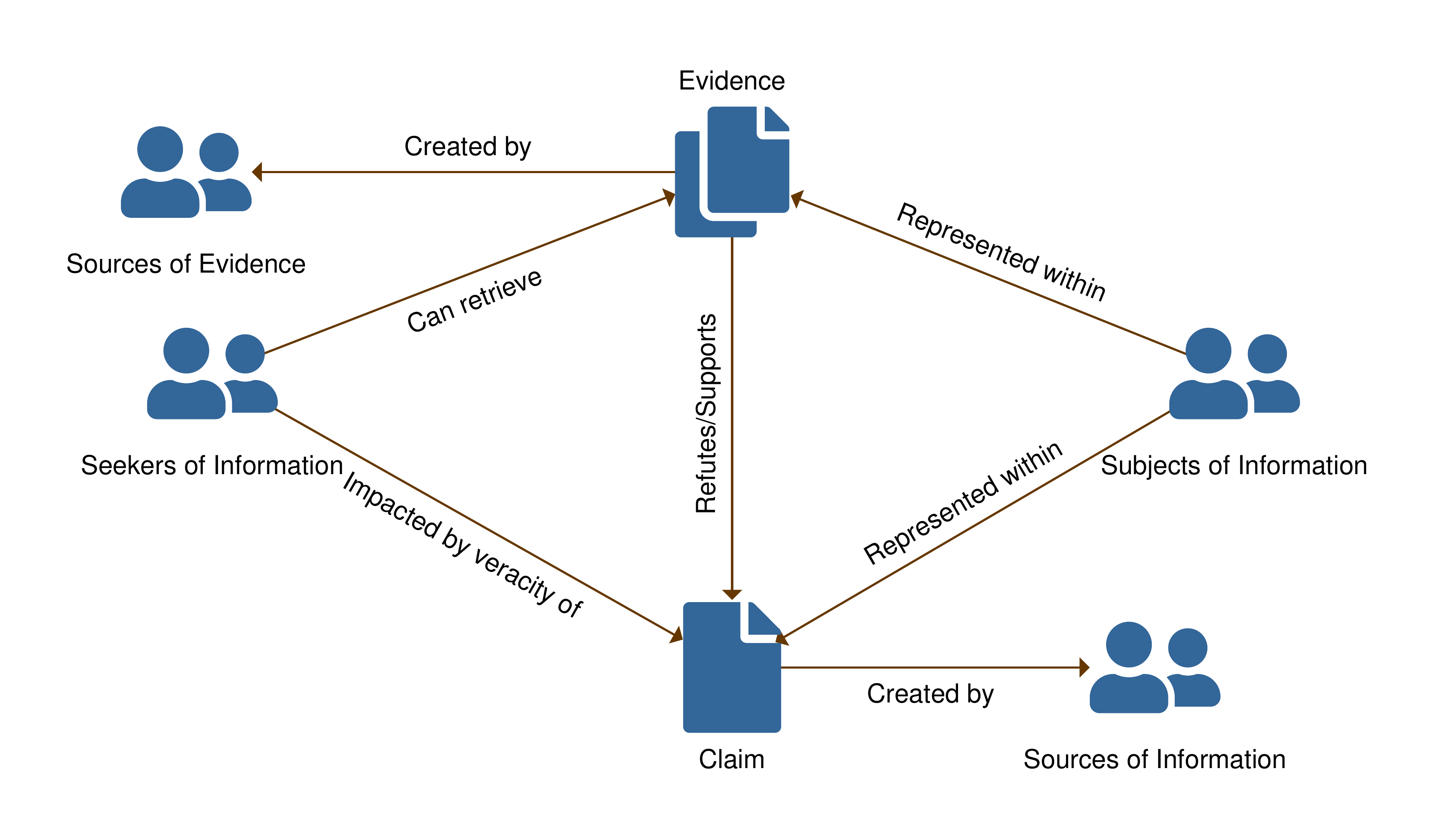}
\caption{The categorisation of fact-checking stakeholders and their graphical relationships, proposed by~\citet{neumann2022}.}
\label{fig:neumann}
\end{figure}

When it comes to the actors of fact-checking, \citet{neumann2022} categorised relevant stakeholders in fact-checking into four groups, as shown in Figure~\ref{fig:neumann}: \emph{seekers of information}, \emph{sources of information}, \emph{subjects of information}, and \emph{sources of evidence}. Seekers of information are those who might be impacted by the veracity of claims and can retrieve evidence that supports or refutes the claims. Examples of them include average users and professionals, e.g., journalists, who use a fact-checking system for veracity assessment. Sources of information are the ones who make the claims to be fact-checked. Authors of articles, content creators generating or posting a claim, and politicians asserting a claim are among the members of this group. Subjects of information, however, comprise individuals, groups, or conceptual topics mentioned in a claim. Finally, sources of evidence refer to individuals generating evidence for fact-checking. These involve domain experts commenting on a relevant claim, users flagging a false claim on social media, and fact-checkers that analyse a suspicious claim.

Moreover, there are a number of studies focusing on exploring false information and fact-checking more systematically. \citet{Shu-K2019b} studied fake news detection and mitigation using network analysis. They proposed a three-dimensional news dissemination ecosystem to describe relationships among different entity types such as publishers, news spreaders, consumers, news pieces, and comments. In their later work of studying disinformation and fake news~\cite{Shu-K2020}, they modelled the `publishing', `spreading', and `social' relationships among publisher, news, and users. \citet{Zhou-X2020} detailed fundamental theories of fake news from various disciplines and presented the fake news life cycle in relation to different entity types including news article, news creator, publisher, and users/spreaders. Similarly, \citet{Mirza-S2023} conducted a series of interviews with disinformation experts of different domains such as fact-checkers, journalists, trust and safety specialists, researchers, and analysts. A disinformation threat framework was proposed based on the insights from the interview to map out threat actors, motivations, capabilities, patterns of attack, attack channels, and targeted audiences. \citet{Zhao-A2025} studied the Chinese social media platform Sina Weibo\footnote{\url{http://weibo.com/}}, with a large-scale dataset of public moderation cases and decision data to investigate the community-driven content moderation system and to examine the three actors (i.e., platform, juror, reporter) involved in the moderation pipelines. They identified inconsistent and contradictory results between the top-down governance (i.e., platform moderator) approach and bottom-up participation (i.e., crowd-sourced moderator) approach. They suggested that there is no single or simple approach to moderate large-scale online content and govern online communities. Finally, \citet{altuncu2025} constructed a dataset comprising nearly 118,000 fact-checking reports from 39 fact-checking organisations accredited by the IFCN and/or EFCSN, and conducted content-driven analyses of the fact-checking ecosystem. Their findings revealed cross-platform differences, including political leanings, which can inform the credibility assessment of fact-checking organisations.

\subsection{Research Gaps}

As mentioned in the previous subsections, a limited number of past studies focused on understanding, analysing and modelling the ecosystem of false information and/or fact-checking. Many of them (e.g., the model shown in Figure~\ref{fig:neumann}, proposed by \citet{neumann2022}) often utilised role-based graphical models to get insights into relationships between different actors and explore the ecosystem. To the best of our knowledge, the use of entity-based graphical models is overlooked, and the role of fact-checking and its impact are less studied in the research literature. The main aim of this paper is to fill this research gap by proposing a conceptual ecosystem for false information and fact-checking using an EER model to discover and represent the comprehensive and complex relationships among different actors (e.g., media outlets, fact-checkers, and false information spreaders) of the false information and fact-checking ecosystem, which could shed light on and potentially encourage researchers/practitioners to model more realistic and complicated cases and facilitate comparison between different studies on different parts or aspects of the ecosystem.

The proposed EER model differs from the previous models in multiple aspects. Firstly, the use of a standard representation, i.e., EER, facilitates the development of computational applications based on the proposed model. Besides, its coverage is comprehensive, with several entities overlooked by previous models, such as regulators, accounts, and associations, allowing the model to offer a more complete view of the ecosystem. Our model also captures the increasing role of automated agents and AI-generated false information in the ecosystem, an area that has received limited attention in ecosystem-level studies.

\section{Modelling the False Information and Fact-checking Ecosystem}
\label{sec:graphical_model}

The ubiquity of smart devices and fast-developing online social media platforms enable everyone to publish, consume, and comment on digital information presented in different formats (i.e., text, image, audio, and video). In addition, traditional media outlets have expanded their services to the internet and online social media platforms on top of their existing channels such as TV, radio, and newspaper, aiming to reach a wider audience. However, such development and revolution make it easier for false information/fake news to spread, but harder to detect and fact-check at scale.

To study the scale and the complexity of the false information and fact-checking ecosystem, we propose an EER model to describe such an ecosystem, aiming to explore the complex relationships between different actors, as well as to empower users with more systematic and comprehensive knowledge and a higher level of conceptualisation to better understand and model real-world scenarios of false information and/or fact-checking. Users of the proposed model are expected to include researchers, practitioners, and a wider range of audiences who would benefit from digital tools developed based on the model, such as journalists, media outlets, and policymakers. The proposed model covers multiple dimensions of the ecosystem, e.g., different fact-checking stakeholders, false information generated, consumed, or verified by the stakeholders, and regulations implemented as a preventive measure.

\subsection{Methodology}
\label{sec:methodology}

Considering the complexity of the false information and fact-checking ecosystem, a systematic approach is needed to construct a comprehensive and reproducible conceptual model of the ecosystem. With this respect, to more systematically understand and model the false information and fact-checking ecosystem, we adopted the `Ontology Development 101' methodology~\citep{Noy-N2001} in this study. Although this study aims to propose an EER model, we preferred to follow an ontology construction methodology to facilitate the proposed model to be utilised for building a new computational ontology as future work. Ontology Development 101 is a simple guide with 7 key phases based on iterative design to provide a basic understanding of ontology development and help developers create an ontology. The rest of this section presents more details of how we adapted the 7 phases to develop the proposed EER model.

\subsubsection{Phase 1: Determine the domain and scope}
\label{sec:phase1}

The first phase of the development process is to determine: 1) the domain of the ontology; 2) the types of questions the ontology should provide answers to; and 3) the potential users of the ontology. The main objective is to gain a comprehensive understanding of the false information and fact-checking ecosystem, including its structure, scale, and interconnections between its actors, and thereby to offer valuable insights into how false information could be generated and spread, as well as the extent to which fact-checking can serve as an effective countermeasure. We would envisage that researchers and practitioners in the relevant domains could be the main users of the proposed model. In addition, software/web tools could be developed based on the model, which could be potentially beneficial to a wider range of audiences including journalists, media outlets, and policymakers.

\subsubsection{Phase 2: Consider reusing existing ontologies}
\label{sec:phase2}

To the best of our knowledge, there is no past work that proposed an EER or other more comprehensive graphical models to study the false information and fact-checking ecosystem despite the existence of simpler role-based models mentioned in Section~\ref{subsec:fifc_ecosystem}, such as the one proposed by \citet{neumann2022}. However, there is some similar work in other domains. Among all, an entity-type graph-based semantic model~\cite{Lu-Y2022} for illustrating the disclosure of users' personal data to different entities and the benefits of such data-sharing activities is an interesting example that inspires us to develop our model presented in this study. In this semantic model for the privacy-benefit trade-off of personal data, there are many entity types (e.g., `Person', `Data', `Service', `Organisation', and `Online group') and a number of edges between entities to describe the complex relationships. It has a similar structure and scale to the false information and fact-checking ecosystem to some extent, where the `Data' entity type can be considered equivalent to the information in the ecosystem, and the privacy-benefit trade-off is similar to the competition between false information spreading and fact-checking activities in the ecosystem. Other entity types in the semantic model can also be mapped to represent other actors of the ecosystem. The next section will present more details of how we determine these important terms as entity types.

\subsubsection{Phase 3: Enumerate important terms}
\label{sec:phase3}

\begin{figure}[!ht]
\centering
\includegraphics[width=0.8\linewidth]{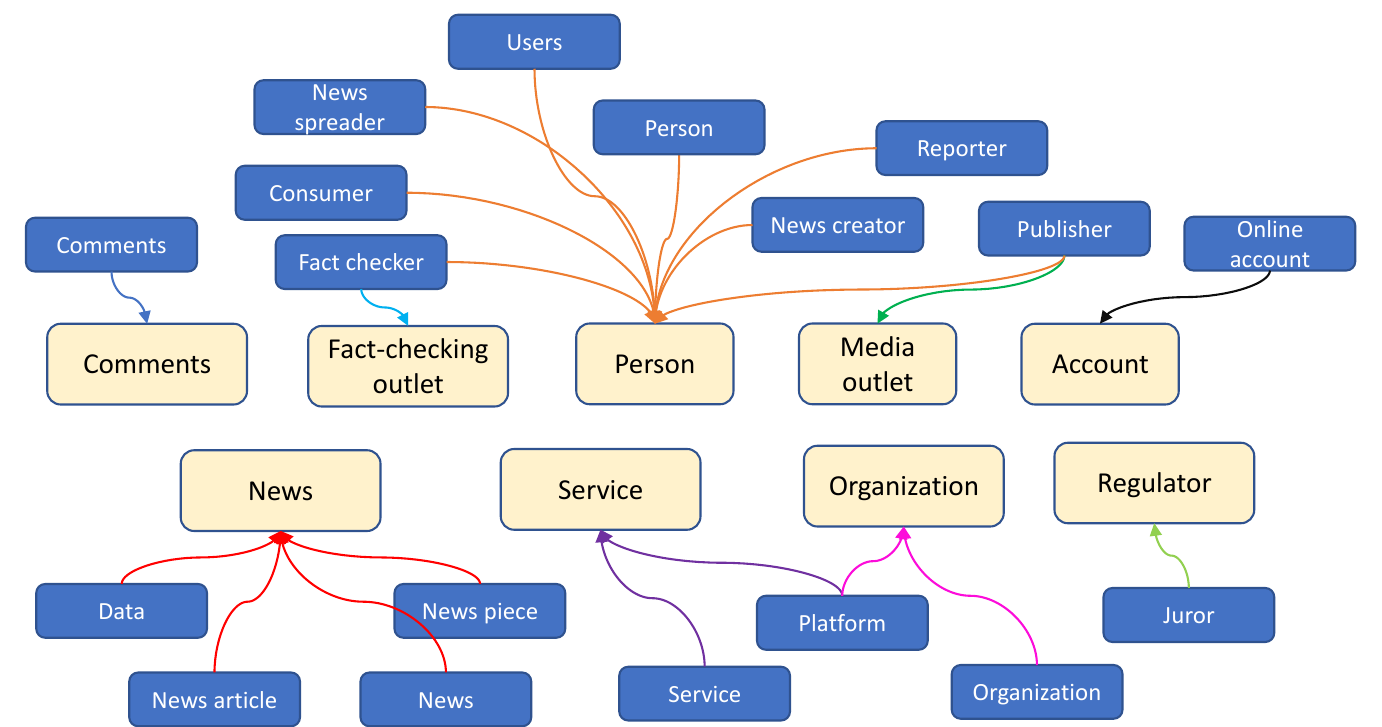}
\caption{Illustration of the process for determining the key terms for the proposed entity-based graphical model, where the entity types in blue boxes are the ones introduced in past work and the entity types in yellow boxes are the high-level entity types we determined for this study.}
\label{fig:phase3}
\end{figure}

This phase of the 101 methodology aims to determine a list of terms used within the ontology with their definition and properties. Initially, our work was inspired by the work that focuses on investigating the false information and fact-checking ecosystem from the perspective of the game theory, where a number of entity types have been identified through different fact-checking scenarios~\cite{Can-B2022}. Then we moved on to review existing similar work to identify key terms. As mentioned in Section~\ref{sec:related_work}, many entity types have been defined in existing literature although they have often been characterised as roles without actually considering what actors can play such roles. By reviewing these entity types and the semantic model of~\citet{Lu-Y2022} mentioned in Section~\ref{sec:phase2}, we tried to categorise and merge the entity types that have a similar definition as illustrated in Figure~\ref{fig:phase3}, and eventually produced the following relevant key terms: `Person', `Media outlet', `Fact-checking outlet', `Service', `News', `Comments', `Account', `Organisation', and `Regulator'. It is worth mentioning that this is not the final selection of entity types for the proposed model. More relevant entity types were derived from the later phases of the 101 methodology. In addition, details of all entity types are presented in Section~\ref{sec:entity_type}.

\subsubsection{Phase 4: Define the classes and the hierarchy}

As shown in Figure~\ref{fig:phase3}, the key term such as `Person' is derived from multiple entity types such as `Users', `Consumers' and `Reporter', that share the similar meaning, suggesting that they could be categorised in the same class and the hierarchy and relationships between them could be also deducted. To further expand the development of classes, a bottom-up approach as recommended by~\citet{Noy-N2001} has been followed. By reviewing and analysing a number of example scenarios (see Section~\ref{sec:real-world_scenarios}) related to the false information and fact-checking ecosystem, we aimed to complete two tasks: 1) to define the hierarchy and relationships for the entity types that have been listed in Section~\ref{sec:phase3}. For instance, we identified that `Person' can belong to a `Media outlet' or a `Fact-checking outlet', `Person' can also create `News' or `Comments'; 2) to discover and identify more closely related entity types with the associated relationships that can supplement the development of the graphical model. For instance, we recognised that `Media outlet' and `Fact-checking outlet' should belong to newly created entity types `Media outlets association' and `Fact-checking association' respectively to adequately reflect the scale of the false information and fact-checking ecosystem. Full details of all entity types and edges can be found in Sections~\ref{sec:entity_type} and \ref{sec:edges}.

\subsubsection{Phases 5 \& 6: Define the properties of classes-slots \& Define the facets of the slots}

Here we merge the two phases of the methodology as they are closely linked. The main aim of Phase 5 is to add necessary properties and information to entity types, and the output of Phase 6 is the ontology (i.e., the EER model in this study). Based on how information can be created and spread in modern society, the developed model covers both the ecosystem in the context of traditional media outlets (e.g., newspaper publishers, TV broadcasters) and user-generated content via the use of digital platforms such as social media services (e.g., X and Facebook) and online video platforms (e.g., YouTube and Vimeo).

\subsubsection{Phase 7: Create instances}

This phase of the methodology is to create individual instances of using a part of the graphical model to see if it can be used for specific domains. We would consider the main aim is to evaluate the effectiveness and flexibility of the proposed model. Thereby, we modelled seven real-world incidents to cover some representative subsets of the overall graphical models, representing sub-ecosystems of the false information and fact-checking ecosystem. More details can be found in Section~\ref{sec:real-world_scenarios}.

\subsection{The Graphical Model of the False Information and Fact-Checking Ecosystem}
\label{sec:model_a}

Following the methodology presented in Section~\ref{sec:methodology}, we developed the graphical model for the false information and fact-checking ecosystem, which could be considered as the foundational abstractions of several other potential formal models. First, it is the graphical depictions of multi-agent systems that are commonly used in AI. Second, it visualises a methodology to approach the issue of fact-checking from a ``knowledge representation and modelling'' perspective. Fact-checking is an interesting ``laboratory'' for knowledge representation and epistemic logic (which is the logic of knowledge) and the current model is a stepping stone to form a full epistemic model of the phenomenon.

\begin{figure*}[!ht]
\centering
\includegraphics[width=\linewidth]{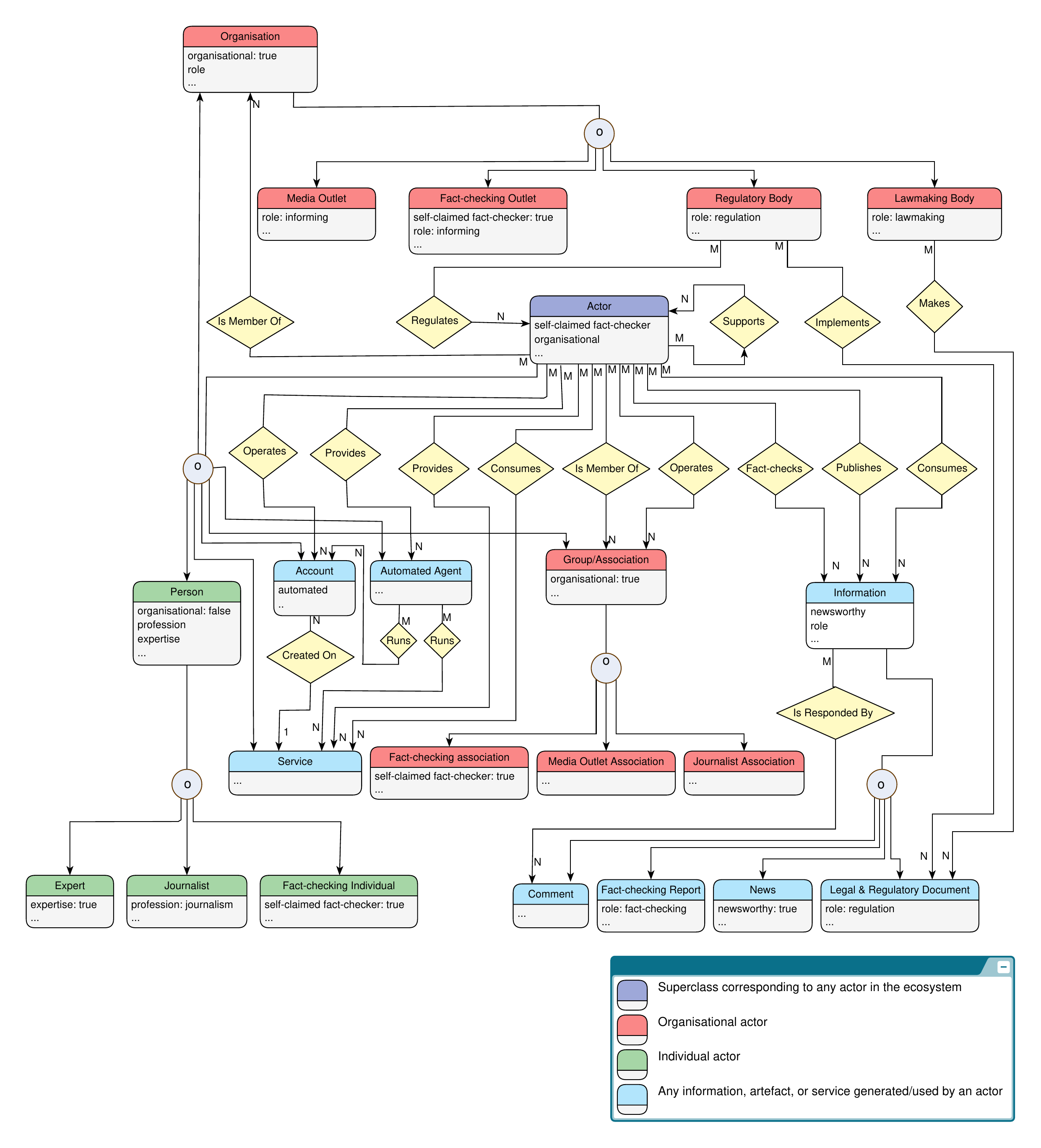}
\caption{False information and fact-checking ecosystem graphical model}
\label{fig:graph_model}
\end{figure*}

The model can be formalised as a directed graph as shown in Figure~\ref{fig:graph_model}, representing different types of entities and their relationships. The graph can be denoted as $G=(V,E)$, where $V=\{V_i\}^N_{i=1}$ and $E=\{E_j\}^M_{j=1}$ represent a set of $N$ nodes and a set of $M$ edges, respectively. Each node $V_i$ represents a type of entity, and each edge $E_j$ describes a semantic relationship between two connected entity types.

The entity types are shown with round-cornered rectangles, while diamond shapes represent semantic relationships between entity types with cardinalities on both sides of the edges. The entity attributes are shown with the labels in the bottom half of the rectangles, ending with three dots to emphasise that more attributes might exist. It is worth noting that the reason to single out these specific attributes is that they are relevant and essential to better describe the context of different semantic relationships between entity types, thereby facilitating explaining and modelling the ecosystem. However, including these attributes does not mean discarding or ignoring other attributes not explicitly shown in Figure~\ref{fig:graph_model}. More different attributes can be added to different entity types to enhance the compatibility and scalability of the graphical model for modelling more specific scenarios under specific contexts. It is also worth noting that we use verbs in different tenses for different textual labels in diamond shapes to capture the importance of time in a semantic relationship.

Entity types with similar definitions are grouped and colour-coded. To simplify the presented diagram by merging some common relationships, a superclass entity covering all the actors in the model, called \emph{Actor}, is added and coloured in indigo. Superclass-subclass hierarchies are shown with the notations adapted from enhanced ER modelling, including the use of small circles with a letter `o' inside implying that the connected subclasses can have common members and `d' implying non-overlapping subclasses. Entity types representing individual actors (i.e., persons), organisational actors, and other entities having relationship(s) with actors are coloured green, red, and blue, respectively. Apart from individual and organisational actors (i.e., green and red rectangles), the \emph{Actor} entity type also encapsulates some entity types shown with blue rectangles that can generate false information, e.g., a social bot belonging to the \emph{Account} entity type (characterised by the attribute \emph{automated} set to \texttt{true}) can create and spread false information online.

\subsubsection{Entity Types}
\label{sec:entity_type}

The proposed model contains $N=21$ different entity types, which are defined below. The entities that can have instances representing individuals or organisations that claim themselves as fact-checkers have an attribute of \emph{self-claimed fact-checker}. In addition, different attributes are defined to distinguish the entity types, including those specific to the subclass entity types and the ones inherited from the superclass, e.g., the \emph{Actor} entity type.

\paragraph{Actor:} An individual, organisation, account, or service involved in the ecosystem of false information and fact-checking. This entity type is added to simplify the proposed model by combining some common relationships that might cause redundancy and connecting to a single entity type. It has six subclasses: 1) \emph{Person}, 2) \emph{Organisation}, 3) \emph{Group/Association}, 4) \emph{Account}, 5) \emph{Automated Agent}, and 6) \emph{Service}.

\begin{enumerate}
\item \emph{Person:} This corresponds to an individual in the physical world. Depending on the role in the ecosystem, it also covers different subclasses shown as separate entity types in Figure~\ref{fig:graph_model}, including \emph{Journalist}, \emph{Expert}, and \emph{Fact-checking Individual}. Besides, there can be many other subclasses, e.g., \emph{Na\"ive User}, \emph{Politician}, or \emph{Legislator}, that are not shown explicitly on the model for the sake of simplicity.
\begin{enumerate}
    \item \emph{Journalist:} Someone who produces and reviews newsworthy content.
    
    \item \emph{Expert:} Someone who possesses the relevant expertise and provides evidence or other help for fact-checking, directly or indirectly and consciously or unknowingly.
    
    \item \emph{Fact-checking Individual:} Someone who drives the fact-checking process and produce fact-checking reports.
\end{enumerate} 

\item \emph{Organisation:} An organised group of people with a particular purpose. While this can have a wide range of subclasses, only the following subclasses are explicitly shown in the model considering their important roles in the false information and fact-checking ecosystem:
\begin{enumerate}
    \item \emph{Media Outlet:} A broadcasting/press channel that can provide information in different formats to the public by different paths, including newspaper, television, radio, and social media. This entity type can overlap with the \emph{Fact-checking outlet} entity type when its \emph{self-claimed fact-checker} attribute (inherited from the \emph{Actor} entity type) is true.

    \item \emph{Fact-checking Outlet:} An organisation dedicated to performing fact-checking or offering fact-checking services, such as Full Fact\footnote{\url{https://fullfact.org/}} in the UK and Snopes\footnote{\url{https://www.snopes.com/}} in the US.

    \item \emph{Lawmaking Body:} A governmental body with the power to make, amend, and repeal laws.

    \item \emph{Regulatory Body:} Any authority having responsibility for the supervision or regulation of any regulated activities or services, e.g., Ofcom\footnote{\url{https://www.ofcom.org.uk/}} in the UK.
\end{enumerate}

\item \emph{Group/Association:} A number of individuals and/or organisations that share common interests, commitments, or purposes. Subclasses of this entity type are as follows: 
\begin{enumerate}
    \item \emph{Media Outlet Association:} An organisation that serves the shared interests of media outlets to protect the general interests of its members in different aspects such as political, regulatory, and legal matters. Examples include News Media Association\footnote{\url{http://www.newsmediauk.org}} in the UK and News Media Alliance\footnote{\url{https://www.newsmediaalliance.org/}} in the US.

    \item \emph{Journalist Association:} A journalism organisation or journalist union that is dedicated to encouraging responsible reporting and ethical behaviour, e.g., British Association of Journalists\footnote{\url{https://bajunion.org.uk/}} and National Union of Journalists\footnote{\url{https://www.nuj.org.uk}} in the UK, and Society of Professional Journalists\footnote{\url{https://www.spj.org/}} in the US.

    \item \emph{Fact-checking Association:} An association that has members of fact-checking outlets that share the same commitment and purposes to promote fact-checking services and duties. Examples of this entity type include IFCN\footnote{\url{https://www.poynter.org/ifcn/}} and EFCSN\footnote{\url{https://efcsn.com/}}.
\end{enumerate}

\item \emph{Account:} A virtual identity of an individual or organisation using online services, such as social media accounts.

\item \emph{Automated Agent:} A software that performs actions autonomously, based on predefined rules, learned models, or algorithmic decision-making, without requiring continuous human intervention. AI agents and content moderation bots are examples of automated agents.

\item \emph{Service:} A platform, tool, or dataset provided by an actor to the public for different uses. This has a wide coverage, including social media platforms, news websites, and services used for fact-checking, e.g., periodic newsletters about recent fact-checks, databases of previously fact-checked articles, and fact-checking tools (e.g., Google Fact Check Tools\footnote{\url{https://toolbox.google.com/factcheck/}}).
\end{enumerate}

\paragraph{Information:} Knowledge or details conveyed in the form of text and/or multimedia and produced/consumed by one or more \emph{Actor}s. The model covers four different types of information -- \emph{News}, \emph{Comment}, \emph{Fact-checking Report}, and \emph{Legal \& Regulatory Document}.
\begin{enumerate}
\item \emph{News:} A report of recent events or information that attracts public interest, published by a media outlet. It can be physical or digital assets in different formats, such as text, image, video, and audio.

\item \emph{Comment:} An actor's opinion, feedback, or other information in response to published news or posted content.

\item \emph{Fact-checking Report:} The outcome of a veracity investigation for a piece of information, which can be published by a fact-checker or a fact-checking outlet in different formats such as a report, a blog, or a news item.

\item \emph{Legal \& Regulatory Document:} A law, or a set of regulations and/or implementation rules made to implement principles of laws to bring them into effect.
\end{enumerate}

\subsubsection{Relationships}
\label{sec:edges}

As previously mentioned, a textual label in a diamond shape indicates a semantic connection between two entity types. It is a somewhat complicated network with a wide variety of relationships between entity types, as shown in Figure~\ref{fig:graph_model}. To better explain the model, it is examined concerning the key roles of the actors through the edges connected to the \emph{Actor} entity type. These roles include information generation and propagation, information consumption, fact-checking, and lawmaking and regulation.

\paragraph{Information Generation and Propagation}
As illustrated in Figure~\ref{fig:graph_model}, information is generated and propagated by (i.e., the \emph{Publishes} relationship) the \emph{Actor} entity type. This can be inherited by its subclass entity types (i.e., all actors in the ecosystem), indicating that a piece of information can be generated by one or more actors. While multiple actors can co-generate information as shown by the many-to-many relationship (M-N notation in the model), the self-referencing relationship \emph{Supports} implies that an actor can also support another actor in information generation in different ways, e.g., financially, by providing know-how, and by sharing personal networks. Besides, the \emph{Operates} relationship between the \emph{Account} and \emph{Actor} entity types implies that those individual and organisational entity types can also have accounts that they use to publish information. More precisely, the proposed model suggests the following four main scenarios for information generation and propagation.

\begin{enumerate}
\item A \emph{Person} can generate information. This covers many cases, such as a journalist publishing a news story, a user commenting on a news article, and a domain expert sharing their views on a subject. Furthermore, a \emph{Person} can use an \emph{Account} to publish user-generated or AI-generated content, e.g., a post shared on a social media platform, a product review left by a consumer on Amazon, or a deepfake image used to manipulate public opinion.

\item An \emph{Organisation}, e.g., a media outlet, a fact-checking outlet or a regulator, can generate information. This can be exemplified by scenarios such as media outlets publishing news, institutions publishing reports, and regulators making statements.

\item Although a \emph{Group/Association} can be considered as an organisation in many cases, it needs a separate connection to the \emph{Actor} entity type since it could generate information without necessarily being an established organisation. An example of such cases is online troll groups coordinated via messaging applications, e.g., WhatsApp and Telegram, organising disinformation campaigns which target particular individuals, organisations, and/or ideologies.

\item An \emph{Automated Agent} can autonomously drive information generation and propagation. For instance, a social media account controlled by a social bot can publish information without any intervention from a human actor. Similarly, an automated online service, such as a search engine or an automated fact-checking tool, can also generate information on its own without human intervention. This also extends to AI agents that may produce false information and/or perform fact-checking tasks.
\end{enumerate}

Apart from the above approaches to generating information, everyone can now readily generate and share user-generated content, such as their stories, videos, and images, thanks to the proliferation of online services, platforms, and service providers that have been made possible by the growth of internet technology. This makes it easier to develop new kinds of media outlets that are distinct from the established ones. The most representative example is \emph{We Media}, i.e., participatory journalism. It was \citet{Bowman-S2003} who originally used this term to describe ``the act of a citizen, or group of citizens, playing an active role in the process of collecting, reporting, analyzing and disseminating news and information.'' \emph{We Media} has become a large and influential media outlet (e.g., accounts of celebrities and social media influencers) in terms of subscribers, and its management still centres mostly on a single person. Although \emph{We Media} is not explicitly modelled here, our model is capable of representing a \emph{We Media} as a \emph{Person}, \emph{Organisation}, or \emph{Group/Association} entity type \emph{Operates} an \emph{Account} entity type, demonstrating the flexibility and scalability of our approach. In addition, such interaction between/among entity types can then be further investigated in order to mimic various motivations (such as monetary, political, and/or religious) of \emph{We Media}, showing the extensibility of the model.

\paragraph{Information Consumption}
The proposed model represents information consumption using the two \emph{Consumes} relationships originating from the \emph{Actor} entity type. While the one connecting to \emph{Information} indicates consumption directly from an information source, the other \emph{Consumes} connecting to \emph{Service} models cases when information is consumed through a service, e.g., reading news using a news aggregator. More precisely, our model supports the following three main information consumption scenarios.

\begin{enumerate}
\item A \emph{Person} can consume information in different ways, such as reading a news item, viewing posts on social media, or studying written material to become trained against false information.

\item An \emph{Organisation} can also consume information, especially for monitoring purposes. For example, a \emph{Fact-checking Outlet} needs to monitor social media posts going viral to determine suspicious claims they need to investigate. As another example, a \emph{Regulatory Body} must monitor information generated by the regulated actor to ensure its compliance with laws and regulations.

\item An \emph{Automated Agent} can consume information through multiple mechanisms, including direct data ingestion from APIs, databases, and knowledge graphs, automated web scraping for data collection, and retrieval-augmented generation (RAG) processes that enable autonomous access to external information sources.
\end{enumerate}

\paragraph{Fact-checking}
In the EER model, fact-checking is covered by the \emph{Actor} entity type through the \emph{Fact-checks} relationship. \emph{Actor} has a specific attribute, \emph{self-claimed fact-checker}, indicating whether an individual or organisational actor declares itself as a fact-checker. This can help identify the actors dedicated to fact-checking in the model.

A \emph{Person} entity type with a \emph{true} value for the \emph{self-claimed fact-checker} attribute denotes that the person considers themselves a fact-checker. Additionally, a \emph{Person} can be a part of a \emph{Fact-checking outlet} modelled via the \emph{Is Member Of} relation between \emph{Actor} and \emph{Organisation}. A \emph{fact-checking report} as a subclass entity type of \emph{Information} can be published as a result of fact-checking (e.g., via the relation \emph{Fact-checks}) by one or more independent fact-checkers (as \emph{Person} entities) or one or more fact-checking outlets (as \emph{Organisation} entities) or a combination of both, where the \emph{self-claimed fact-checker} attribute is set to be \emph{true} by default for both entity types.

In addition to fact-checkers and fact-checking outlets, media outlets also inherit the \emph{self-claimed fact-checker} attribute and can perform fact-checking via specific units/programs dedicated to fact-checking, such as \emph{FactCheck}\footnote{\url{https://www.channel4.com/news/factcheck}} of Channel 4 News and \emph{BBC Verify}\footnote{\url{https://www.bbc.com/news/reality_check}} (formerly BBC Reality Check). Such fact-checking services often have an advantage over independent fact-checkers and fact-checking outlets as they can reach wider audiences~\citep{Graves-L2016}.

It is important to note that the \emph{Actor} entity type is connected to the \emph{Service} entity type via \emph{Provides} and \emph{Consumes} relationships, signifying that \emph{Person}, \emph{Group/Association}, and \emph{Organisation} entities can 1) produce various services in different formats, including platforms, tools, and datasets to share with the fact-checking community; and 2) use such services to cross-reference, fact-check, and do additional analysis. Besides, the \emph{Automated Agent} entity type can also perform fact-checking autonomously, for example, in the form of an AI agent trained as a fully automated fact-checking system.

\paragraph{Lawmaking and Regulation}
Regulators must issue and implement different regulations to enforce laws issued by lawmaking bodies in various circumstances. These are covered by the relationships between the \emph{Lawmaking Body} and \emph{Legal \& Regulatory Document} entity types, as well as the \emph{Regulatory Body} and \emph{Legal \& Regulatory Document} entity types. The \emph{Lawmaking Body} entity type is responsible for issuing laws necessary (i.e., modelled via the relationship \emph{Makes}) to regulate the ecosystem. The \emph{Regulatory Body} entity type, however, regulates (i.e., modelled using the relationship \emph{Regulates}) the actors involved in the ecosystem, such as the \emph{Person}, \emph{Organisation}, and \emph{Group/Association} entity types, by implementing (i.e., modelled via the relationship \emph{Implements} the issued laws and regulations, i.e., legal directives on how to implement laws. For instance, the UK's Ofcom\footnote{\url{https://www.ofcom.org.uk/}} (i.e., the \emph{Regulatory Body} entity type in the model) regulates communication services, including fact-checking outlets (i.e., the \emph{Fact-checking Outlet} entity type), and its powers and procedures in the context of fact-checking are outlined in the UK's Online Safety Act 2023 (i.e., the \emph{Legal \& Regulatory Document} entity type).

\subsection{Additional remarks}
\label{sec:remarks}

As shown in Figure~\ref{fig:graph_model}, an entity type can have the same semantic relationships with multiple entity types. For instance, any \emph{Actor}, including a \emph{Media Outlet}, a \emph{Fact-checking Outlet}, and a \emph{Person} can fact-check the same \emph{News} item. In such a situation, competitive or collaborative fact-checking relationships from multiple entity types will normally be present. Getting more insights about such phenomena would require further investigation of the other entity types with their edges connecting to them. Such a snowballing effect can lead to interesting observations and findings that would require the use of other theories such as game theory to offer better elaboration. One example would be the game between regulators, news outlets, and fact-checkers, which would consider not just financial benefits and reputation gains, but also legal costs. Note that regulators may also act illegally and can pay legal costs if their wrong decisions are challenged by media outlets or fact-checkers.

\section{Example Application 1: Modelling Real-World Scenarios}
\label{sec:real-world_scenarios}

In this section, we show how the proposed graphical model can be used to model relevant real-world scenarios as well as reveal insights that are often not immediately apparent. All diagrams of the seven presented example real-world scenarios are produced following the same colour scheme used for the graphical model as depicted in Figure~\ref{fig:graph_model}. The example scenarios include both real-life incidents and hypothetical scenarios, and they are selected to cover a representative subset of the overall graphical model, representing a sub-ecosystem.

\subsection{BBC Breakfast Incident}
\label{sec:bbc}

\begin{figure*}[!htb]
\centering
\includegraphics[width=\linewidth]{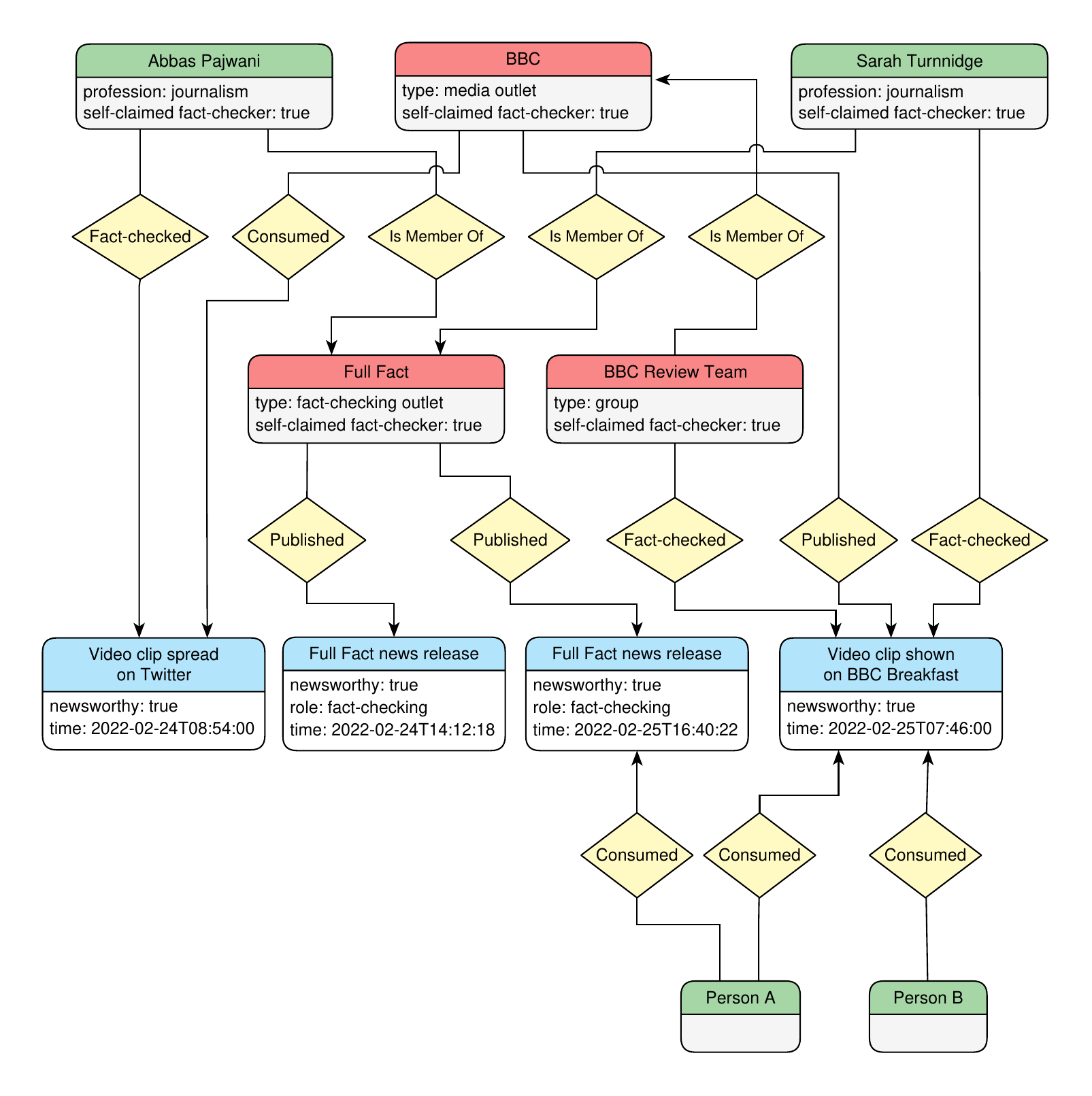}
\caption{Modelling a real-world scenario that involves BBC and a number of fact checkers}
\label{fig:graph_model_bbc}
\end{figure*}

On 25th February 2022, \textit{BBC Breakfast} played a clip of military planes and claimed that they were Russian military flying into Ukraine following the invasion of Ukraine~\citep{Turnidge-S2022}. As shown in Figure~\ref{fig:graph_model_bbc}, \textit{Sarah Turnidge} from \textit{Full Fact}, who is also a journalist, conducted fact-checking on this video clip and prepared a fact-check report, which was later published as a news release by \textit{Full Fact}. The report stated that the footage from BBC Breakfast was preparations for a military parade near Moscow from May 2020, but not about the invasion as it claimed. As depicted in Figure~\ref{fig:graph_model_bbc}, from the audience's perspective, \textit{Person B} is more likely to spread the misinformation than \textit{Person A} who first watched the \textit{BBC Breakfast} show, and then, later read about the fact-check report released by \textit{FullFact}. One interesting finding as illustrated in the figure is that the video clip broadcast on BBC Breakfast Show was originally from a widespread video clip on Twitter, which was also fact-checked by other fact-checking outlets such as the following: 1) On 24th Feb 2022, \textit{Maldita} fact-checked the video clip on Twitter and published a news release on its website~\citep{Maldita2022}; and 2) On 24th Feb 2022, another journalist, also a fact-checker \textit{Abbas Panjwani} from \textit{Full Fact} conducted fact-checking on this video clip, and a corresponding news release was published by \textit{Full Fact}~\citep{Panjwani-A2022}.

Despite all the fact-checking reports being priorly available, such misinformation was still in the wild. Based on Sarah Tunidge's fact-checking report~\citep{Turnidge-S2022}, one spokesperson from BBC confirmed that ``The footage was used once briefly in error and the team have been reminded about verifying images.'' This suggests that BBC have an internal review team equipped with a certain level of internal review or a fact-checking mechanism as demonstrated in Figure~\ref{fig:graph_model_bbc}, yet the video clip with the false claim was broadcast. This demonstrates that the use of the proposed model can analyse real-world incidents and help researchers/practitioners to systematically uncover its real scale and scope. For example, the \emph{time} attributes of the blue entities in Figure~\ref{fig:graph_model_bbc} can help a chronological analysis of the incident. Moreover, if we further extend this sub-graph to bring in other entity types and semantic relationships, more analyses can be done to investigate this incident further. Finally, the presented model of the incident exemplifies how the EER model can be utilised to construct a data model for a knowledge graph of existing fact-checking reports, which can facilitate the detection of such incidents in future. 

\subsection{A Person Using Fact-checking Services and Resources}
\label{sec:ifcn}

\begin{figure*}[!htb]
\centering
\includegraphics[width=\linewidth]{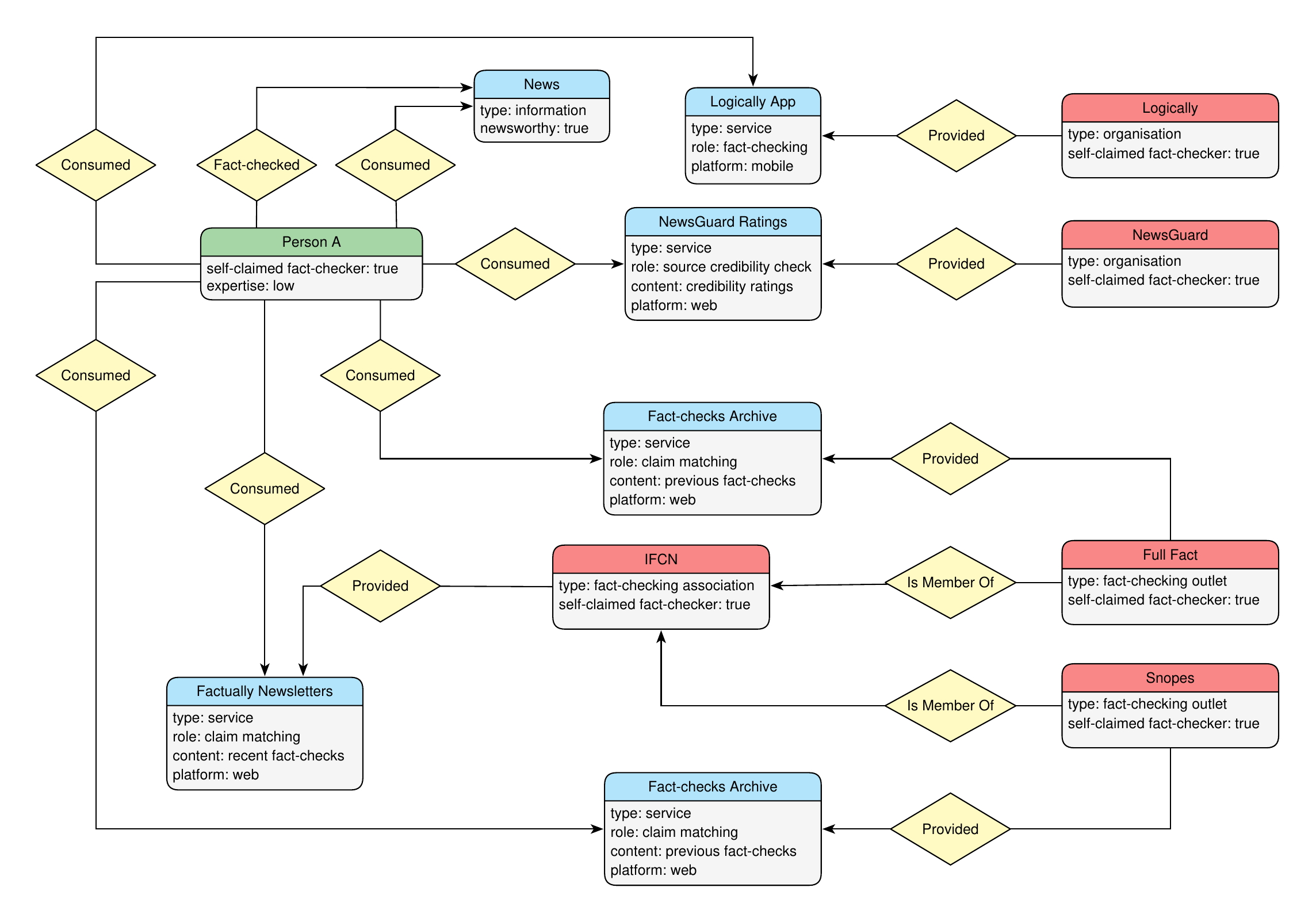}
\caption{Modelling fact-checking services and resources provided by fact-checking outlets}
\label{fig:graph_model_services}
\end{figure*}

This scenario includes modelling fact-checking services and resources provided by fact-checking organisations, companies, associations, and networks by using the proposed model. Figure~\ref{fig:graph_model_services} exemplifies the scenario with multiple services and resources that a person with low expertise in fact-checking can benefit from to perform fact-checking independently. Given a news article including a number of suspicious claims, \textit{Person A} examines if the claims have been previously fact-checked by any fact-checking organisation, e.g., \textit{Full Fact}, and \textit{Snopes}, through their fact-checks archives where \textit{Person A} can search for previous fact-checks. Besides, \textit{Person A} can review the recent fact-checks via periodical watch newsletters published by IFCN, i.e., \textit{Factually Newsletters}, which contains the recent fact-checks collected from the verified signatories of IFCN. Nevertheless, \textit{Person A} can also subscribe to the newsletters published by the IFCN-verified fact-checking organisations directly to receive more country-specific fact-checks. In addition, \textit{Person A} can also publish comments in correspondence to the news article, as well as to read comments published by others.

Apart from fact-checking organisations and networks, many companies provide solutions that others can make use of while fact-checking. If the suspicious claims have never been fact-checked by a fact-checking organisation, \textit{Person A} requests verification from Logically's fact-checking team through the \textit{Logically App\footnote{\url{https://www.logically.ai/products/app}}}. Other than the content of the news article, \textit{Person A} can assess the credibility of the news source by checking the source's \textit{NewsGuard Rating\footnote{\url{https://www.newsguardtech.com/solutions/newsguard/}}}, which is the trust rating published by NewsGuard. Here, the model illustrates the complex landscape of the fact-checking ecosystem, highlighting the interactions between various actors and showcasing the possible paths a person can take to fact-check news using various resources. This can help identify a customised set of fact-checking services and resources for different target audiences based on a number of factors, such as their expertise, accessibility requirements, age group, or even, trust in different actors in the ecosystem.

\subsection{Voluntary Regulators of the UK Media}

In this scenario, we intend to demonstrate that the voluntary regulators sub-ecosystem of the model proposed in Section~\ref{sec:graphical_model} can be explored and enhanced to reveal the complexity of the sub-ecosystem. In this scenario, we list some representative media outlets including \textit{Telegraph}, \textit{Daily Mail}, \textit{The Mail}, \textit{The Financial Times}, \textit{The Independent}, and \textit{The Guardian}.

\begin{figure*}[!htb]
\centering
\includegraphics[width=\linewidth]{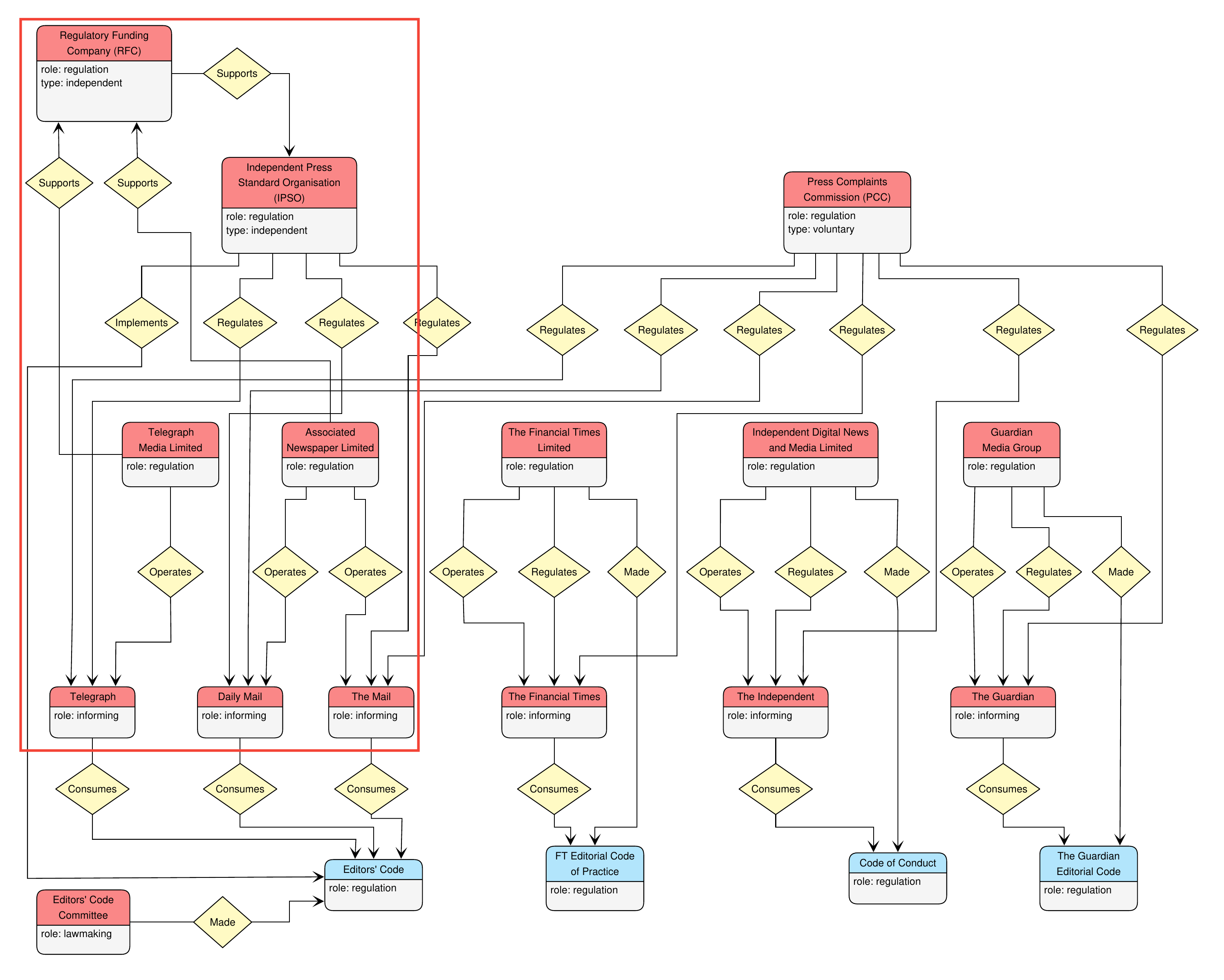}
\caption{Modelling independent regulators that regulate the media outlets in the UK. The big red rectangle indicates a conflict of interest between multiple entities.}
\label{fig:graph_model_regulator}
\end{figure*}

There are many active independent regulators (which can be considered a sub-type of the \emph{Regulatory Body} entity type) in the UK. Since 2014, The \textit{Independent Press Standard Organisation (IPSO)} has served as the voluntary regulator for British printed newspapers and magazines, implementing the \textit{Editor Code} created by \textit{Editor Code Committee}, continuing regulating \textit{Telegraph}, \textit{Daily Mail}, and \textit{The Mail}. As shown in Figure~\ref{fig:graph_model_regulator}, \textit{IPSO} is financed (i.e., the \emph{supports} relationship) by \textit{Regulatory Funding Company (RFC)}, which is funded by \textit{Telegraph Media Limited} and \textit{Associated Newspaper Limited}. It is worth noting that, \textit{Telegraph Media Limited} owns \textit{Telegraph} and \textit{Associated Newspaper Limited} owns \textit{Daily Mail} and \textit{The Mail}. Such relationships raised some concerns, hence \textit{The Financial Times}, \textit{The Independent}, and \textit{The Guardian} refused to join \textit{IPSO}. Instead, each chose to publish their own standards/guidance to self-regulate as illustrated in Figure~\ref{fig:graph_model_regulator}. It is not our position to judge how fact-checking ecosystem should be regulated, however, such critical insights into conflicts of interest can be easily identified using the proposed model, demonstrating its utility and capability in aiding researchers and practitioners to conduct in-depth analysis. Furthermore, a more complete model of existing regulators using the proposed model can assist policymakers in identifying possible legal gaps to be covered by new laws and regulations.

\subsection{Journalists and Their Associations in the US}

\begin{figure*}[!htb]
\centering
\includegraphics[width=\linewidth]{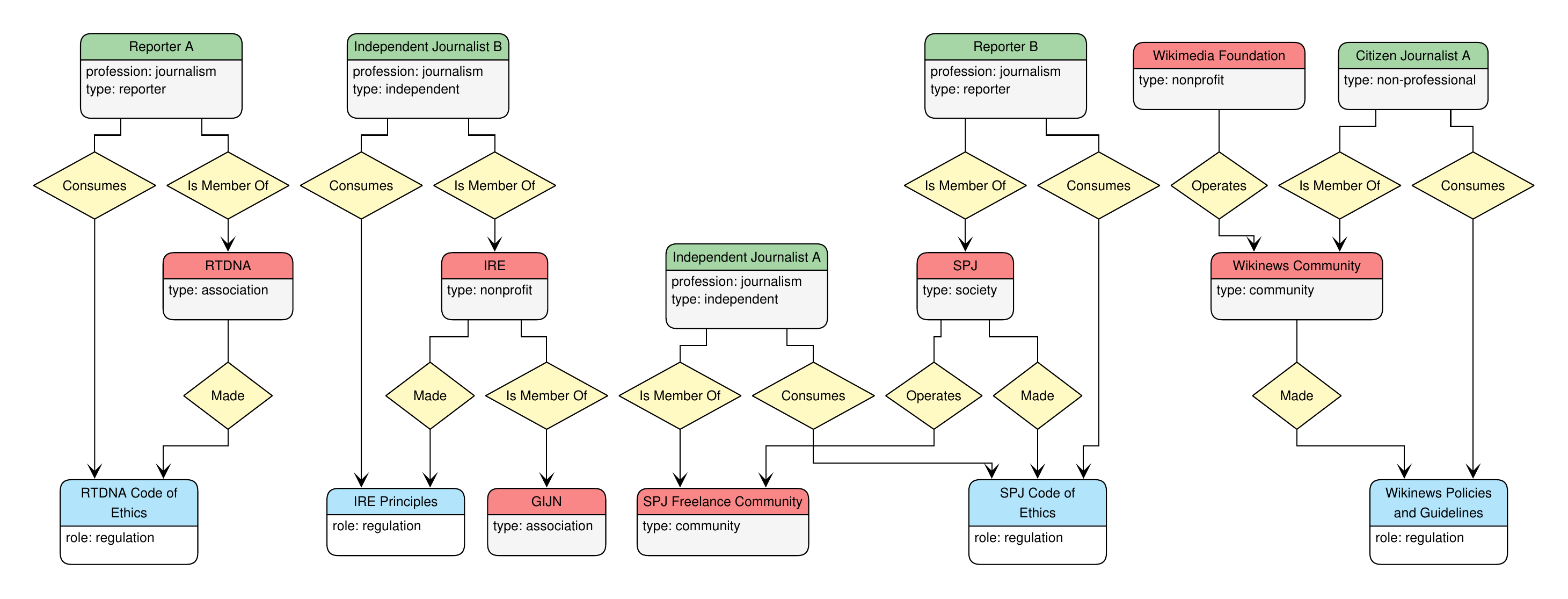}
\caption{Modelling different journalist types and journalist associations}
\label{fig:graph_model_journalist}
\end{figure*}

In this scenario, we aim to demonstrate the relationships between different types of journalists and their associations with the proposed graphical model. Such a demonstration can be used to examine what principles journalists are required (or suggested) to comply with, according to their associations. This can facilitate the process of source credibility assessment for the published news as well as the evaluation of journalists based on to what extent they comply with the principles. As shown in Figure~\ref{fig:graph_model_journalist}, for the journalist associations on a national scale, we preferred to cover a number of US-based associations as an example.

For journalists working for a media outlet in the US, i.e., reporters in Figure~\ref{fig:graph_model_journalist}, The Society of Professional Journalists (SPJ)\footnote{\url{https://www.spj.org/}} and The Radio Television Digital News Association (RTDNA)\footnote{\url{https://www.rtdna.org/}} are two associations representing journalists. Both associations provide their \textit{Code of Ethics} as a guideline for their members to follow while performing their work. Besides, they provide several tools and resources to support the journalism activities of their members.

Another group of journalists are \textit{independent journalists}, i.e., freelance journalists, which are professional journalists that do not work for any media outlet, but publish news on their own (e.g., via social media) or in collaboration with other media outlets. For independent journalists, there also exists some communities and organisations that provide support for their journalism activities. SPJ has a community dedicated to independent journalists, called \textit{SPJ Freelance Community}\footnote{\url{https://www.spj.org/freelance.asp}}, that offers tools and resources to help their members, as well as organises events and programs. Another example is \textit{Investigative Reporters and Editors Inc.\ (IRE)}\footnote{\url{https://www.ire.org/}}, which is a grassroots not-for-profit organisation dedicated to supporting investigative journalism. IRE supports independent journalists through the \textit{Freelance Investigative Reporters and Editors (FIRE)} project\footnote{\url{https://www.firenewsroom.org/}}, which is a service bureau providing grants and editorial services for freelance journalists. IRE is a member of \textit{Global Investigative Journalism Network (GIJN)}\footnote{\url{https://gijn.org/}}, and all IRE members must follow the \textit{IRE Principles}.

Finally, citizen journalists are another type of journalists, where individuals publish news collaboratively. \textit{Wikinews}\footnote{\url{https://www.wikinews.org/}} is a project of \textit{Wikimedia Foundation}, that serves as a platform for citizen journalism, where users can collaboratively publish news stories provided that they follow the \textit{Wikinews Policies and Guidelines}.

\subsection{Trump's Twitter Account Suspension Incident}

\begin{figure*}[!ht]
\centering
\includegraphics[width=\linewidth]{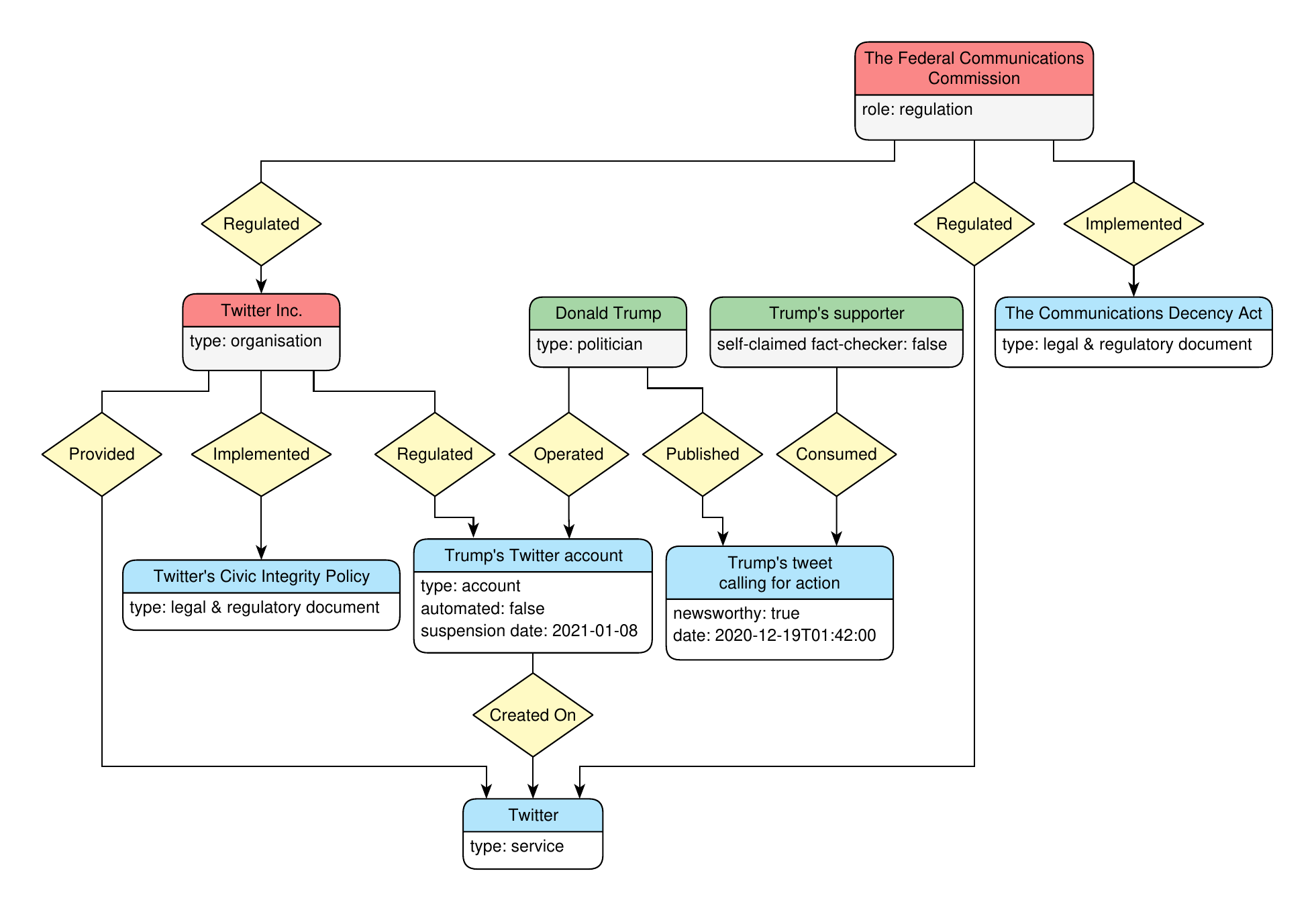}
\caption{A graphic model explaining the US Capitol riot in 2021}
\label{fig:graph_model_dt_twitter}
\end{figure*}

This scenario is based on a real incident in 2021, in which Twitter (now X) permanently suspended the account of the then ex-President of the US, Donald Trump. Figure~\ref{fig:graph_model_dt_twitter} shows the graphical model of the relationships between different actors involved in the incident using our proposed model. After losing the 2020 US presidential election, Trump posted multiple tweets containing some false claims that the presidential election had been stolen and calling his supporters to action~\citep{poynter2021capitol}. Consequently, Trump's supporters gathered and attacked the US Capitol, which caused the death of five people. As a result, on 8 January 2021, Twitter Inc.\ permanently suspended Trump's Twitter account due to ``the risk of further incitement of violence'', based on an assessment of Trump's tweets under its \emph{Civic Integrity Policy}~\citep{twitter2021suspension}. And, the US communications regulator, the \emph{Federal Communications Commission (FCC)}, did not object to Twitter's decision~\citep{reuters2021fcc}. The graphical model reveals that content published by Donald Trump and consumed by a modelled supporter of him was not assessed or confirmed (i.e., there is no \emph{fact-checked} relationship in the model), highlighting the lack of a fact-checking mechanism on Twitter at that time and the need for better content moderation. Weeks after the US Capitol riot on January 6 2021, Twitter launched Birdwatch (now, Community Notes), a crowdsourcing-based content moderation program, to debunk misinformation and propaganda~\citep{twitter2021birdwatch}.

\subsection{Fact-checking Multimedia Content}

\begin{figure*}[!htb]
\centering
\includegraphics[width=\linewidth]{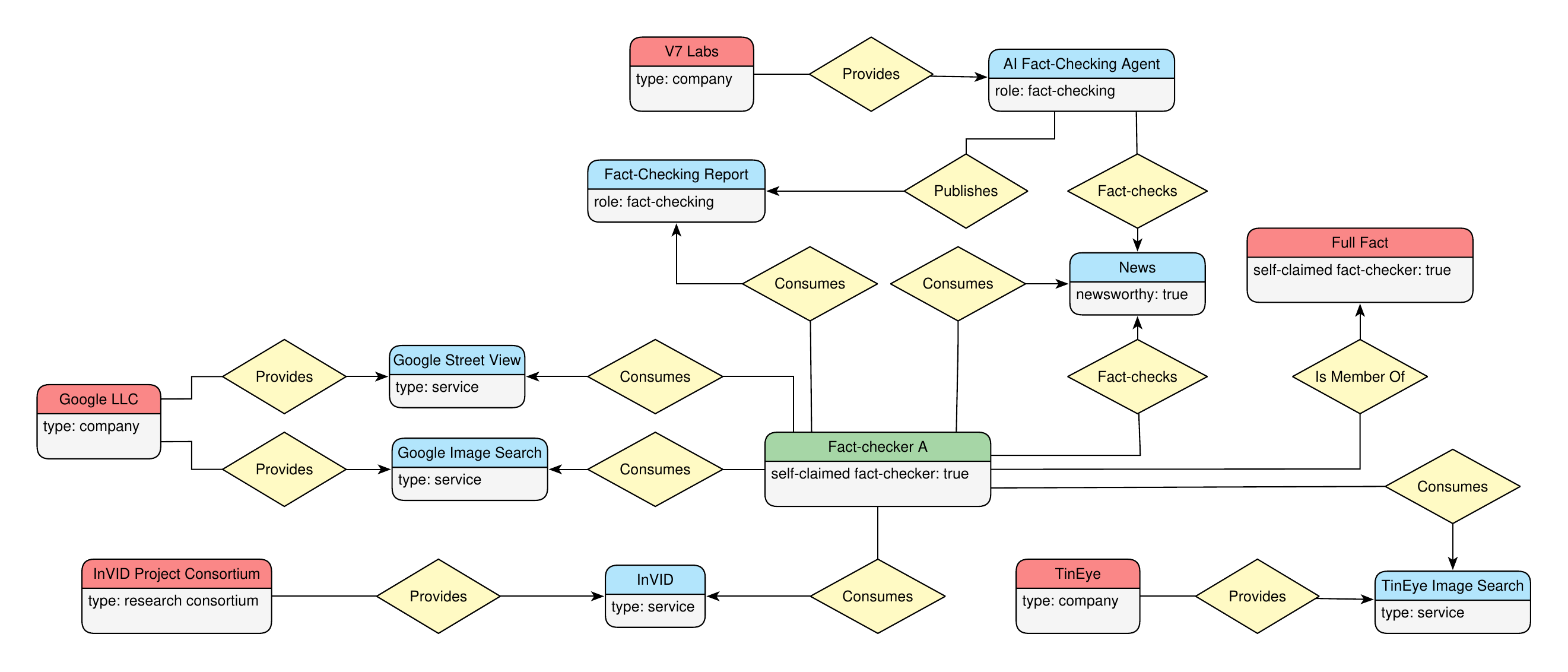}
\caption{A graphic model of how fact-checking platforms investigate the veracity of multimedia content}
\label{fig:graph_model_services_ugc}
\end{figure*}

False information based on fake, doctored, or misleading images and videos is quite common in UGC platforms as it is more likely to attract users' attention. In this manner, this scenario explores how fact-checking platforms investigate the veracity of multimedia content. Based on multiple tutorials on fact-checking multimedia content, published by fact-checking platforms~\citep{rahman2022, krishna2020, wilksharper2018}, Figure~\ref{fig:graph_model_services_ugc} models the fact-checking process performed by fact-checking platforms with the help of multiple tools and services. Fact-checkers utilise reverse image search tools, such as Google Image Search and TinEye\footnote{\url{https://tineye.com/}}, to identify the source of a particular image and whether it is taken out of its context. Furthermore, geolocating the image with detailed mapping tools, e.g., Google Street View, can be useful to obtain more clues about the context of the image. Finally, an example AI agent developed for fact-checking\footnote{\url{https://www.v7labs.com/agents/fact-checking-agent}} is incorporated into the model to illustrate the potential role of AI agents in delegating and automating specific tasks, thereby supporting and streamlining standard fact-checking workflows. The example scenario demonstrates the possibility of utilising our model to establish a knowledge base and guidance on how to use various resources to conduct multimedia content fact-checking tasks, this could be useful for fact-checkers or anyone interested in fact-checking or conducting related research. We envisage that the further expansion of such a graphical model can be beneficial to both fact-checking researchers and practitioners.

\subsection{Fact-checking Fact-checkers}

\begin{figure*}[!htb]
\centering
\includegraphics[width=\linewidth]{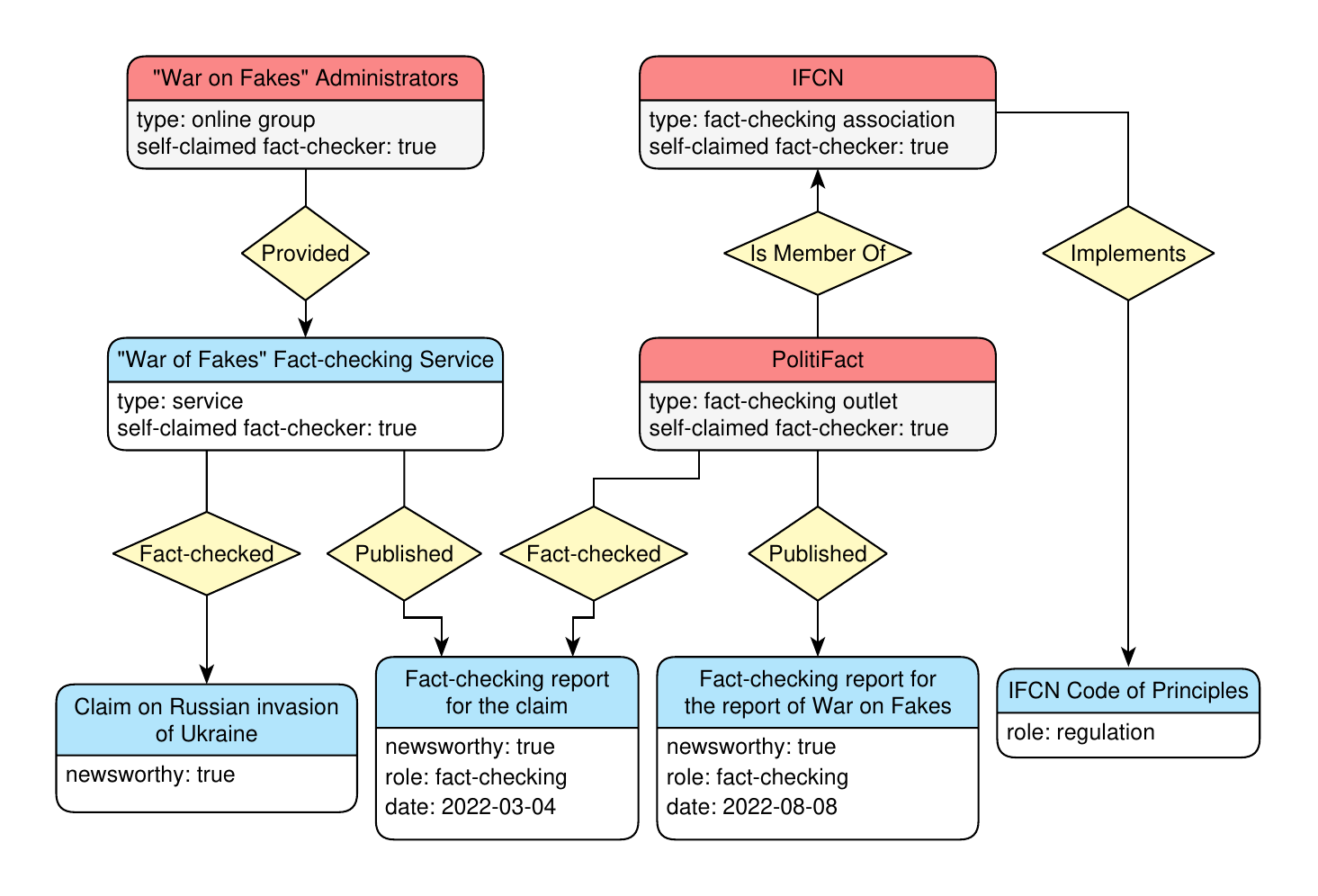}
\caption{Modelling a real-world incident involving PolitiFact fact-checking a fact-checking report}
\label{fig:factchecking_factcheckers}
\end{figure*}

This scenario is based on an actual incident during the Russian invasion of Ukraine. Shortly after the beginning of the invasion, a newly launched pro-Russian ``fact-checking service'', namely \emph{War on Fakes}, published a piece for debunking the claim that Ukrainians were not propagating disinformation against Russians. However, an IFCN-verified fact-checking platform, \emph{PolitiFact} reviewed over 380 ``fact-checks'' published by \emph{War on Fakes} in English and judged their so-called ``fact-checks'' to be Russian disinformation~\citep{romero2022}. Figure~\ref{fig:factchecking_factcheckers} corresponds to the graphical model based on the incident. Our model can be leveraged to distinguish fact-checking outlets following widely acknowledged standards and principles (e.g., IFCN Code of Principles) from self-proclaimed fact-checking groups or organisations that aim to make propaganda, using the \emph{Is Member Of} relationship of the corresponding entity. This scenario shows that fact-checking is a dynamic process, which might occasionally require fact-checking fact-checkers.

\section{Example Application 2: Reviewing Research Literature}
\label{sec:literature_analysis}

We have demonstrated some capabilities of the proposed model using example scenarios listed in Section~\ref{sec:real-world_scenarios}. In addition, we would like to illustrate how such conceptual models can be used to conduct more fine-grained analyses of the research literature, with a more systematic categorisation of previously studied topics based on the entities and relationships represented by our model. This can also assist researchers in more systematically identifying potential research gaps related to specific parts of the ecosystem. With this respect, this section aims to demonstrate how the proposed model can be utilised to facilitate the literature analysis for a number of relevant publications recently published. 

\subsection{Methodology}

For the selection of the relevant publications, we performed a search in Google Scholar with the following search query and limited the search results to those published in 2023. Note that the year 2023 was chosen as an example, as our aim here is to demonstrate how a literature review can benefit from our proposed model. 

\texttt{\small "misinformation" OR "disinformation" OR "false information" OR "fake news" OR "fact-checking"}

The top 10 peer-reviewed English articles were selected from the search results after being sorted by relevance. However, one publication that showed up in the search results was excluded since it appeared only because it used an included keyword in an irrelevant context. A wide range of academic fields, including Communication Studies~\citep{vandermeer2023, freiling2023, hameleers2023}, Psychology~\citep{altay2023, lawson2023, modirrousta2023}, Computer Science~\citep{roy2023}, Marketing~\citep{dholakia2023}, Economics~\citep{azzimonti2023}, and Political Science~\citep{erlich2023}, were represented in the papers that were ultimately chosen. All selected papers were qualitatively analysed after the paper selection step to find topics relevant to the entity types and relationships in the proposed model and then encoded with incrementally specified tags in the format ``$\text{Entity}_1\text{RelationshipEntity}_2$''. Such a systematical generation of tags can facilitate the tagging process of typical literature reviews. Table~\ref{tab:literature_analysis_tags} shows the list of tags with the corresponding papers, identified in consequence of the analysis. These tags were then used to review the literature in terms of different aspects of the ecosystem by grouping the papers containing discussions on the same or similar relationships covered by the model.

\begin{table}[!htb]
\caption{The list of generated tags and relevant papers for each tag}
\centering
\begin{tabularx}{\linewidth}{cX}
\toprule
\textbf{Tag} & \textbf{Relevant Paper(s)}\\ 
\midrule
UserConsumesInformation & \cite{freiling2023, dholakia2023, azzimonti2023, vandermeer2023, modirrousta2023}\\
UserPublishesInformation & \cite{freiling2023, lawson2023}\\ 
UserFactchecksInformation & \cite{erlich2023, altay2023, roy2023}\\
GroupPublishesInformation & \cite{lawson2023}\\
OrganisationPublishesInformation & \cite{hameleers2023}\\
MediaOutletPublishesInformation & \cite{dholakia2023}\\
AccountPublishesInformation & \cite{azzimonti2023}\\
ExpertProvidesService & \cite{roy2023, modirrousta2023}\\
RegulatorRegulatesService & \cite{altay2023, hameleers2023, dholakia2023}\\
\bottomrule
\end{tabularx}
\label{tab:literature_analysis_tags}
\end{table}

\subsection{Literature Analysis}

In general terms, the tags shown in Table~\ref{tab:literature_analysis_tags} indicate that user-information relationships were more studied than other portions of the graphical model. Besides, most studies focused on information generation or consumption, rather than fact-checking. Here we summarise the three main themes that have been discussed in the literature we reviewed.

\subsubsection{False information sharing}

More precisely, the literature involves attempts to better understand the factors and motivations behind believing and sharing false information. \citet{freiling2023} found that political beliefs, anxiety, and their interactions can affect whether people believe and share false claims. Moreover, \citet{lawson2023} illuminated the role of conformity pressure and social costs within online groups people belong to when it comes to false information dissemination. Apart from being ideologically and socially motivated, \citet{hameleers2023} pointed out that false information can also be disseminated in the form of disinformation with political motivations such as disinformation campaigns originating from the Russian Internet Research Agency (IRA).

As examined by \citet{dholakia2023}, the current economic organisation of media outlets also contribute to the generation and dissemination of false information. With the emergence of social media platforms, user data has become another profitable product for media outlets. Such platforms offer a `phatic communication' environment to their users, lacking the structural mechanisms and editorial processes of the news media outlets that can avoid false information. In addition, bots on social media platforms are among the spreaders of false information~\citep{azzimonti2023}.

\subsubsection{Debunking and prebunking}

When it comes to dealing with false information, recent research also covers different debunking and prebunking strategies and their consequences. As \citet{erlich2023} demonstrated with their study conducted with Ukrainians, people are capable of distinguishing false information from true information in many cases. They follow various strategies to identify false information, from cross-checking and following recent fact-checks to different digital tools such as search engines and online encyclopaedias. However, \citet{altay2023} pointed out that more research is needed on the uses and misuse of such digital tools. In addition, there are efforts to develop fully automated systems to detect false information. For instance, \citet{roy2023} proposed a deep stacked Long short-term memory (LSTM) model for false information prediction on social platforms to identify it as soon as posted on the social platform.

As remarked by \citet{vandermeer2023}, attempts to combat false information, such as fact-checking and misinformation warnings, can also have the unintended consequence of a decrease in general trust in factual correct information. Therefore, it is crucial not only to fight against false information but also to reestablish trust in legitimate news. \citet{modirrousta2023} came to a similar conclusion in their overview of gamified inoculation interventions which aim to increase people's resistance against false information by exposing them to common false information generation techniques. Their findings suggest that such interventions did not improve discrimination between true and false news, but increased false responses to all news items.

\subsubsection{False information Regulation}

Finally, the collected papers discussed the existing ecosystem in terms of its regulation. With the digital media ecosystem, including social media, traditional gatekeepers no longer control the media landscape, facilitating the dissemination of false information alongside true information, according to \citet{altay2023}. Therefore, as \citet{hameleers2023} underlined, social media platforms and big tech companies need to play a central role in regulating this digital ecosystem to protect their users from false information online. With this respect, \citet{dholakia2023} suggested the adoption of new legal frameworks toward holding social media networks accountable for limiting the spread of false information on their platforms, based on the success of legal frameworks concerning privacy protection.

\subsection{Summary}

By integrating systematic tag generation with the proposed model, we can swiftly identify key themes in the literature, providing a clear understanding of current research trends. This approach not only highlights key topics but is also capable of categorising papers with similar research focus. While this can be achieved by existing methods, such as topic modelling, they are less systematic and are more likely to miss important topics studied in the literature. Moreover, it reveals potential research areas that have been overlooked, thus pointing to future research directions within the scope of false information and the fact-checking ecosystem.

\section{Further Discussions and Future Work}
\label{sec:discussion}

The proposed graphical model can be used for studying false information online and fact-checking in many different ways. In this section, we discuss some of the possible applications and possible future research work.

One obvious extension of the proposed graphical model is to define a computational ontology and automatically build such an ontology from different (closed and public) data sources. Doing so is not trivial since automatically discovering entities and relationships between different entities can be challenging. It is likely advanced techniques including NLP (natural language processing) techniques such as named entity recognition (NER) algorithms, supervised machine learning methods and large language models (LLMs) will be necessary. In addition, the leverage of a computational ontology can formalise the proposed model, facilitating large-scale Internet measurement studies to better understand how false information spreads and how different actors behave, and allowing automated reasoning to automatically discover new phenomena related to false information and fact-checking. A computational ontology derived from the proposed graphical model can also be incorporated into existing fact-checking tools to make them more aware of the graphical nature of the ecosystem. For instance, a fact-checking tool could leverage the computational ontology to identify the most critical nodes and paths in order to identify the best communication strategy.

As demonstrated in Section~\ref{sec:real-world_scenarios}, the graphical model can be used to guide a more systematic categorisation of real-world scenarios of false information and fact-checking, and how the different scenarios are related to each other. Such categorisation can help researchers, practitioners, and policymakers better understand the complexity of the false information and fact-checking ecosystem, identify better interventions, and evaluate such interventions. Modelling real-world scenarios can be improved in future by automating the modelling and visualisation processes. This can be achieved by leveraging automated knowledge graph construction methods~\citep{zhong2023}. Besides, different scenarios can be compared automatically to facilitate more in-depth analysis.

The example of utilising the proposed model for literature analysis sheds light on its usefulness and novel support for guiding literature review on related topics. A potential future work for this application might be automating model-based literature analysis, which was performed manually in the given example. Moreover, the graphical model can also help guide the design of empirical studies, e.g., in identifying relevant stakeholders so that the recruitment of human participants can be more representative.

Another important area of applications of the graphical model is modelling and computer-based simulation of the false information and fact-checking ecosystem, especially agent-based models, which can directly borrow the entities and relationships in the graphical model. Finally but not least, the many different types of conflicting and cooperative relationships between different entities in different false information and fact-checking scenarios can also help researchers to use various theoretical tools such as game theory and epistemic logic to rigorously study the spread of false information, changes of behaviours of different entities, and also the overall evolution of such ecosystems.

\section{Conclusion}
\label{sec:conclusion}

In this paper, we report the first graphical model of the ecosystem around false information and fact-checking, with an EER model, in order to conceptually capture the complicated relationships between many different entities. The usefulness of such a conceptual model is demonstrated using some example real-world scenarios and an example literature review, and its wider applications for studying false information online and fact-checking are discussed. It is our hope that the work can stimulate more work on conceptual modelling of false information online and fact-checking.

\subsubsection*{Funding}

No funding was received for conducting this study.

\subsubsection*{Data Availability Statement}

No datasets were generated or analysed during the current study.

\bibliography{main}

\end{document}